\documentclass[10pt,letterpaper]{article}
\usepackage{opex3}

\begin{document}


\title{A Common-Path Interferometer for Time-Resolved and Shot-Noise-Limited
Detection of Single Nanoparticles}

\author{Meindert A. van Dijk, Markus Lippitz$^*$, Dani{\"e}l Stolwijk,\\and Michel Orrit}
\address{MoNOS, Huygens Laboratory, Universiteit Leiden, P.O.
Box 9504, 2300 RA Leiden, The Netherlands\\
$^*$ present address: Institut f{\"u}r Physikalische Chemie,
Johannes Gutenberg-Universit{\"a}t, Welderweg 11, D-55099 Mainz,
Germany}

\
\email{orrit@molphys.leidenuniv.nl} 

\homepage{http://www.monos.leidenuniv.nl/} 


\begin{abstract}

We give a detailed description of a novel method for time-resolved
experiments on single non-luminescent nanoparticles. The method is
based on the combination of pump-probe spectroscopy and a
common-path interferometer. In our interferometer, probe and
reference arms are separated in time and polarization by a
birefringent crystal. The interferometer, fully described by an
analytical model, allows us to separately detect the real and
imaginary contributions to the signal.  We demonstrate the
possibilities of the setup by time-resolved detection of single
gold nanoparticles as small as 10 nm in diameter, and of acoustic
oscillations of particles larger than 40~nm in diameter.
\end{abstract}






\section{Introduction}

For the last 15 years, single-molecule spectroscopy has mainly
focused on luminescence studies of fluorescent molecules and
semiconductor nanocrystals \cite{science99}. In the last few
years, however, ever increasing interest in the properties and
applications of metal nanoparticles (nanospheres, nanorods, etc.,
with sizes between 1 and 100~nm) has stimulated the development of
various optical detection schemes for single metal nanoparticles.
Two recent articles have reviewed the general far-field optical
methods for the detection of single nanoparticles
\cite{vandijk05jul}, and more particularly the absorption and
scattering-based methods \cite{vandijkpccp}. An attractive
characteristic of gold nanoparticles is their high photostability.
Unlike dyes or semiconductor nanocrystals, gold particles do not
suffer from photobleaching or photoblinking, which makes them
appealing labels for biophysicists \cite{cognet03sep30}. Another
strong motivation is the study of original physical properties of
metal nanoparticles, that often differ from those of the bulk
metal, and to a large extent can be tuned through the size and
shape of the nanoparticle. This sensitivity to size and shape,
however, makes studies of ensembles of nanoparticles particularly
vulnerable to distributions in sizes, shapes, crystal defects,
etc.. Isolating a single particle once and for all eliminates
inhomogeneous broadening and any implicit averaging inherent to
even the most carefully selected ensembles. Only single-particle
experiments permit to study a particle's elastic interaction with
its specific close environment \cite{vandijk05dec31}, to correlate
optical and structural properties \cite{jin05sep14}, or to obtain
new insight in their linear and nonlinear optical properties
\cite{berciaud05mar,muskens06mar,lippitz05apr}.

The number of available methods for the detection of single metal
nanoparticles has grown rapidly in the last few years
\cite{vandijk05jul}. Of all these methods, interferometric
detection of the scattered field seems very promising, mainly
because the interferometric signal drops as the third power of the
particle size only, whereas the direct scattering signal drops as
the sixth power, but also for the possibility to detect the full
complex response of a nanoparticle
\cite{vandijk05dec31,stoller06aug15}. In all interferometric
experiments, the field scattered by a single metal nanoparticle is
mixed with a reference field, but the methods differ in their
choices for the reference and scattered field \cite{vandijkpccp}.
Using the incident wave itself as a reference, one basically
measures absorption \cite{arbouet04sep17}. Alternatively, the
reference wave can be the reflection from the substrate on which
the particle has been deposited \cite{lindfors04jul16}, or an
external reflection in a Michelson interferometer
\cite{ignatovich06jan13}. The signal wave can be either directly
scattered by the particle, or indirectly scattered from a local
inhomogeneity of refractive index, for example induced by heat
released by the excited particle into its surroundings, as done in
the photothermal method \cite{boyer02aug16,berciaud04dec17}. In a
recent work, we have used a common-path interferometer to measure
the time-resolved response of single gold nanoparticles down to a
10-nm diameter \cite{vandijk05dec31}. This sensitive technique
allowed us to detect acoustic vibrations of single gold
nanoparticles. The present paper presents a more detailed
description of our interferometer and its full characterization.

In our common-path interferometer, probe and reference arms are
spatially overlapped (in contrast to, for example, Michelson and
Mach-Zehnder interferometers \cite{kop95dec}). The interfering
waves are distinguished on the basis of both polarization and
time, although any single one of these characteristics would in
principle suffice. Temporal separations can be created via a
relative delay between laser pulses, as done here, or via a small
phase shift between two CW beams, as for example in phase-contrast
microscopy \cite{zernike42}, or by counter propagation in a Sagnac
ring interferometer, as was recently implemented by Hwang et al.
\cite{hwang06feb}, for the phase-sensitive detection of very weak
absorbers. The first implementation of a common-path
interferometer in combination with pulsed laser sources was
published by LaGasse et al. \cite{lagasse89mar15} in an optical
switching experiment. In this experiment, the probe pulse was
delayed by means of polarizing beam splitters and mirrors. Patel
et al. \cite{patel98may10} have built a number of logical elements
by using the same technique, but with a birefringent fiber as a
delay medium. Hurley and Wright \cite{hurley99sep15} constructed a
common-path interferometer based on a Sagnac interferometer, also
using diversion through polarizing beam splitters to split a pulse
into probe and reference. Polarization analysis allowed them to
separately detect pump-induced changes in reflectivity and phase.
Using this method, the same group were able to image the
progression of pump-induced acoustic waves along a surface
\cite{sugawara02may6,tachizaki06apr}.

We also rely on a polarization-induced delay between a probe pulse and a
reference pulse, but we use a birefringent crystal as beamsplitter. The main
advantage of this element is its simplicity and ease of alignment, which leads
to a high contrast of the interference pattern. The delay between the
interfering wavetrains in combination with a pump-probe configuration allows us
to obtain information on ultrafast properties of the nanoparticles, and the high
sensitivity and low noise floor enables experiments on single nanoparticles. By
choosing the proper configuration for the polarization optics of our
interferometer, we can separate the detection of amplitude and phase changes
induced by a single nanoparticle on the probe pulse, and measure the full
complex response of the particle. Although we used our interferometer only for
gold nanoparticles so far, the method is in principle
applicable to any absorbing and/or dispersing nano-object.

The paper is organized as follows. Section~\ref{sec:method} describes the
experimental method. Section~\ref{sec:model} provides a model description of the
setup based on Jones algebra. In section~\ref{sec:results} we discuss a few
experiments that were performed with the setup. In particular, we characterize
the setup by comparing data to the mathematical model
(subsection~\ref{secsub:charac}), we perform a noise analysis showing that our
setup is detection-limited but very close to the photon-noise limit
(subsection~\ref{secsub:noise}), and we provide some examples of measurements on
single gold nanoparticles
(subsection~\ref{secsub:gold}).

\section{Experimental Method}
\label{sec:method}

\begin{figure}[htp]
\begin{center}
\includegraphics[width=110mm]{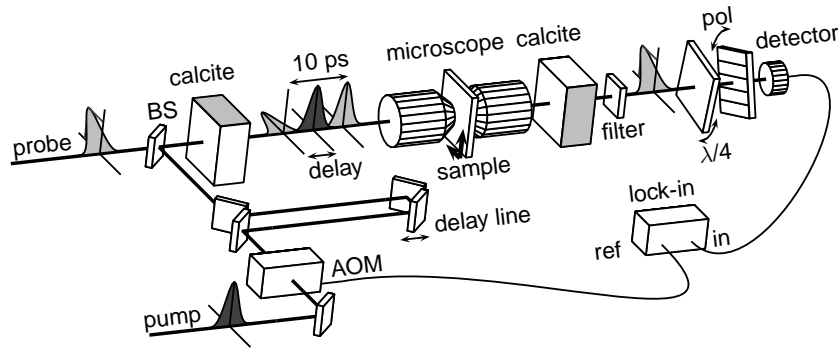}
\end{center}
\caption{\label{setup}Schematic drawing of the pump-probe
common-path interferometer. A pump pulse  and a pair of reference
and probe pulses are focused on the sample in a microscope. The
reference-probe pulses arise from a single pulse, split in time
(10~ps delay) and polarization by a properly oriented birefringent
calcite crystal. The delay between the pump pulse and the
reference-probe pair can be scanned with a delay line. After the
microscope, probe and reference are recombined by a second crystal
and their interference monitors pump-induced changes in the
optical properties of the sample. A quarter-wave plate
($\lambda/4$) and a polarizer (pol) are independently rotated to
set the working point of the interferometer. The other details are described in the text.}
\end{figure}

We have designed a polarization-based common-path interferometer that uses two
birefringent crystals to split and recombine the probe and reference waves
\cite{vandijk05dec31}. A sketch of the setup is shown in Fig.~\ref{setup}. A
laser pulse that is initially polarized at 45 degrees from vertical is split
into two orthogonal polarization directions by the first calcite crystal, whose
fast optical axis is vertical. Propagation in the thick (15~mm) birefringent
crystal delays one of the polarizations by a time longer than the pulse length.
This effectively creates two pulses, a probe and a reference, which are
orthogonally polarized and have a mutual delay of 10~ps. These pulses interact
with the sample (a glass coverslide on which nanoparticles are spincoated)
in the microscope.

Behind the microscope, probe and reference are recombined by a second crystal,
identical to the first one but oriented perpendicularly, its fast axis now being
horizontal. With an analyzer assembly of a quarter-wave plate and a polarizer,
the recombined waves are projected onto the same polarization state and
interfere. Independently rotating the quarter-wave plate and the polarizer
allows us to tune the relative contributions of the phase and amplitude of the
probe and reference waves. The interference reports on changes of the sample
over the time interval between the probe and the reference pulses. In order to
generate and accumulate the signal, we induce a change of the sample by means of
a separate laser pulse, in a pump-probe configuration. This pump pulse (from a
synchronous second laser source) is intensity-modulated by an acoustic-optical modulator
(AOM) and travels over a variable delay line, before it is overlapped by a dichroic beam splitter (BS) with the
probe beam. Dichroic filters efficiently prevent pump light
from reaching the detector. By varying the delay between the pump pulse and the
pair of probe and reference pulses, we can measure the
ultrafast dynamics of our sample.

Let us schematically describe the working of the ideal
interferometer with proper settings of the analyzer assembly. If
the sample does neither modify the phase nor the amplitude of the
probe with respect to those of the reference (e.g., if the pump
laser is switched off, or if there is no particle in the laser
focus), the polarization state after recombination is again
linear, rotated by 45 degrees from the vertical. If we now switch
on the pump laser while a nanoparticle sits in the common focus of
the two laser beams, a small change in the phase and/or amplitude
of the probe pulse will modify the polarization state after
recombination, giving rise to a slightly elliptical and/or rotated
linear polarization. This changes the transmission through the
analyzer assembly. The change in polarization state is thus
translated into a change of intensity at the detector, which we
can measure with a lock-in amplifier synchronized with the pump's
modulation.

With the quarter-wave plate and the polarizer, we can tune the
fraction of the light that reaches the detector. If the
quarter-wave plate has one of its optical axes parallel to the
polarization direction of the incoming light, and if the polarizer
is perpendicular to the incoming polarization, the detected
intensity is minimized, and the interferometer is in a dark
fringe. By rotating the polarizer by 90 degrees, we find a bright
fringe, where the detected intensity has a maximum. Besides tuning
the fraction of probe light that reaches the detector, we can also
use the quarter-wave plate and the polarizer to select a working
point where either phase or amplitude modifications of the probe
field by the sample can be detected, independently from each
other. If  quarter-wave plate and polarizer are rotated in such a
way that the  amplitudes of the interfering fields are equal, but
the phases are different, the interference term will only report
on phase changes of the probe -- assuming small changes to the
fields as it is the case with nanoparticles. Conversely, if the
polarizer and the quarter-wave plate are tuned so that the
amplitudes are different while the phases are the same, the
interferometer only senses amplitude changes of the probe, while
phase changes remain unnoticed. The precise orientations of the
polarizer and the quarter-wave plate for which the interferometer
is either purely amplitude-sensitive or purely phase-sensitive are
calculated in section \ref{sec:model} using a model based on Jones
matrices.

Key features of our interferometer are the birefringent crystals
for temporal beam splitting and recombination. Although they have
the disadvantage that the delay between the probe and the
reference pulse is fixed, the alignment becomes much easier, since
it is not necessary to align the overlap of the spatial paths of
the two interferometer arms. This considerably improves the
visibility of the interferometric fringes. The crystals are 15-mm
thick calcite crystals, cut from the same slab to make them as
alike as possible. In order to keep the operation of the
interferometer stable, we had to actively stabilize the
temperature difference between the two crystals. No special
temperature or mechanical stabilization is needed for the other
elements in the beam path as they  affect probe and reference in
exactly the same manner.

If there is a small difference between the optical path lengths
through the crystals, either because of a difference in
orientation, temperature, or because of a difference in thickness
(the crystals are specified to be equally thick to within
1~$\mu$m), we will have to allow for a static change in the
polarization state of the probe beam after recombination. Instead
of being linear at 45 degrees, the recombined beam can be
elliptically polarized with a rotated axis, even when the pump is
off. A second static effect that must be taken into account is a
residual difference in the transmissions of the vertical and
horizontal polarizations by the microscope. These two effects
change the orientations of the quarter-wave plate and the
polarizer at the dark fringe, and the settings of the
phase-sensitive and amplitude-sensitive working points. These
experimental imperfections, however, can be included into the
model of section \ref{sec:model}. By measuring them, we can
determine and compensate for their influence on the response of
the interferometer. The quality of an interferometer can be
characterized by its contrast ratio $I_{max}/I_{min}$. It can be
shown that the previous effects, if properly compensated for, do
not reduce the contrast. The contrast ratio is in fact limited by
other imperfections of the optical elements, such as
depolarization by the microscope objectives.

To generate our probe beam, we start with the signal beam of an
OPO (APE, Berlin) synchronously pumped by a Ti:Sapphire laser
(Coherent Mira 900D) at a repetition rate of 76~MHz. The signal
beam is frequency-doubled in the OPO cavity. The resulting pulses
are nearly Fourier-limited and are tunable in wavelength between
520 and 650~nm. The Ti:Sapphire laser is operated at 800~nm, and
pumped by a CW frequency-doubled Nd:YAG laser (Coherent Verdi V10,
10~W, 532~nm). The laser system can be aligned to produce pulse
lengths either in the picosecond or in the femtosecond range. The
pulse lengths (full width at half maximum of the autocorrelation
function) of the Ti:Sapphire laser and the OPO are approximately
3~ps and 1.5~ps in the picosecond configuration and 120~fs and
250~fs in the femtosecond configuration. Unless otherwise
indicated, the measurements in this article were carried out in
the femtosecond configuration.

A small part of the Ti:Sapphire laser beam, split off before
entering the OPO, serves as the pump in the experiment. The pump
beam travels over a variable delay stage to tune its time of
arrival with respect to that of the probe-reference pair (see
Fig.~\ref{setup}). The pump beam is combined with the probe beam
by a dichroic beam splitter before the first crystal. The pump
polarization is kept parallel to one of the optical axes of the
crystal, to ensure that this pulse is not split. After the
objectives, the residual pump light is filtered out by optical
bandpass filters.

The home-built microscope consists of two objectives, an
oil-immersion objective (NA 1.4) for excitation and an air-spaced
objective (NA 0.95) for collection. The sample is mounted on a
piezo-electric stage and can be scanned with 25-nm precision.
Commercial gold nanoparticles (purchased from British Biocell
International and Sigma-Aldrich and used without further
treatment) were dispersed in a 10~mg/ml aqueous solution of
polyvinyl alcohol and the resulting suspension was spin-coated
onto a clean glass microscope cover slide.

Our detector is an analog Silicon APD (Hamamatsu C5331-11), which
is sensitive for signals modulated at frequencies between 4~kHz
and 100~MHz. The smallest optical signal that can be detected
against electronic noise, called the noise-equivalent power (NEP),
is specified between 0.5 and 1~$\mathrm{pW/\sqrt{Hz}}$. The
detector signal is fed into a lock-in amplifier (Stanford SR844),
whose internal clock drives an acousto-optical modulator (AOM) in
the pump path at 400~kHz. To prevent overload of the lock-in from
the 76~MHz signal of the laser pulses, the signal from the
detector is pre-filtered with a passive 12~MHz low-pass filter.

\section{Model of the interferometer}
\label{sec:model}

We model our interferometer by representing each element by a
Jones matrix. In this way, we obtain analytical expressions for
the intensity at the detector and for the measured signal, as
functions of the angles of the quarter-wave plate $\phi$ and of
the polarizer $\theta$ (see Fig.~\ref{setup}). We calculate the
contribution to our signal of real and imaginary changes in the
electric field of the probe for each $\theta$ and $\phi$ and
calculate for which combination of angles we can measure the real
and imaginary changes separately.

From Fig.~\ref{setup}, we  write all elements of the
interferometer as Jones matrices. Matrix multiplication of the
elements with the incoming probe field will yield the electric
field and subsequently the intensity at the detector.

The input wave is polarized at $-45$ degrees, $E_{in} = \sqrt{1/2}
\, (1,-1)$, and is split into a horizontally polarized reference
wave, which travels in front, and a vertically polarized probe.
The absorption of the pump pulse by the particle triggers a small
time-varying modification $\zeta(t)$ in the fields transmitted
though the microscope. The modification $\zeta$ is complex-valued,
allowing for changes in amplitude and phase of the fields. The
temporal zero is the arrival of the pump pulse at the sample.
Reference and probe pulses are orthogonally polarized on each
other and delayed by 10~ps. 
After recombination, the Jones matrix associated with the particle
can be written as
\begin{equation}\label{eq:pump1}
    \mathbf{P} = \left( \begin{array}{cc} 1 + \zeta_H(t - 10 \textrm{ps}) & 0 \\ 0 & 1+
{\zeta'}_V(t) \\ \end{array} \right) \quad .
\end{equation}
Shape and orientation of the particle are \textit{a priori}
unknown, and a difference in particle response in the horizontal
and vertical direction has to be taken into account, and is here
indicated with the indices $H$ and $V$. Also, the field of the
reference pulse might alter the state of the particle, which
changes the particle response to the probe pulse from $\zeta(t)$
to $\zeta'(t)$. Including these two effects would however lead to
more parameters than are solvable, so we assume for now that the
particle is symmetric for the horizontal and vertical direction
(so either spherical, or spheroid with the two equal axes in the
x-y plane), and that we are in a linear-response regime, where the
state of the particle is not affected by the field of the
reference pulse.
We can now write, instead of
Eq.~(\ref{eq:pump1}),
\begin{equation}\label{eq:pump}
    \mathbf{P} = \left( \begin{array}{cc} 1 + \zeta(t - 10~\textrm{ps}) & 0 \\ 0 & 1+
\zeta(t) \\ \end{array} \right)  \; \approx \;  \left( \begin{array}{cc} 1  & 0 \\ 0 & 1+
\Delta\zeta(t) \\ \end{array} \right) \quad ,
\end{equation}
where $\Delta\zeta(t) = \zeta(t) - \zeta(t - 10~\textrm{ps})$ is
the 10-ps difference in the pump-induced modification of the
field. The approximation is exact for $t < 10~\textrm{ps}$ as then
$\zeta(t-10~\textrm{ps}) = 0$,  and introduces only small errors
for working points of the interferometer close to the dark fringe
for all other delays $t$.


As was already mentioned in section \ref{sec:method}, it is
important to consider a difference in optical path length through
both crystals and a difference in transmission of probe and
reference through the microscope. The former effect manifests
itself as an additional wave plate in the optical path, while the
latter can be modelled with a relative decrease in amplitude of
one of the two arms. In the model, we can combine both effects in
one matrix $\mathbf{D}$, which adds two additional parameters,
$\rho$ for the difference in transmission between the two
polarizations (dichroism), and $\tau$ for the optical-path
difference:
\begin{equation}
\label{eq:calib}
  \mathbf{D}(\rho,\tau)=
  \frac{1}{1+(1-\rho)^2}\left( \begin{array}{cc}  1 & 0 \\ 0 & (1-\rho)\ e^{
i\tau} \\ \end{array} \right) \quad .
\end{equation}
The prefactor is introduced for normalization of the bright fringe
intensity. $\rho$ and $\tau$ can be determined experimentally,
since a change in either of them causes the dark fringe of the
interferometer to deviate from its original position
$(\theta,\phi)=(45^\circ,45^\circ)$. Therefore, by measuring the
actual position of the dark fringe, the two calibration parameters
$\rho$ and $\tau$ can be calculated.

\begin{figure}[tbp]
\begin{center}
\includegraphics[width=120mm]{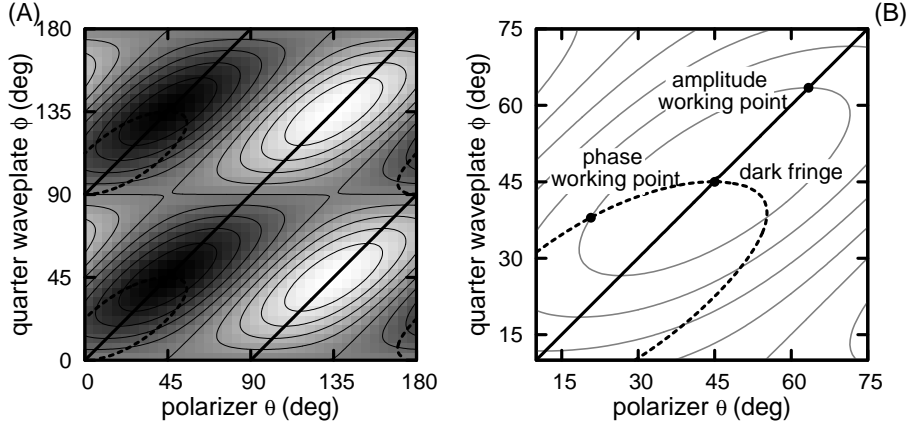}
\end{center}
\caption{\label{wps_theory} (A) Contour plot of the intensity at
the detector as a function of the angle of the quarter-wave plate
and the polarizer, showing bright and dark fringes. On the thick
lines, the interferometer is sensitive to either amplitude changes
only (solid) or phase changes only (dashed). (B)  Magnification
around a dark fringe, additionally showing two working points of
the interferometer where to detect pure amplitude or phase
signals. }
\end{figure}

The Jones matrices of a quarter-wave plate $\mathbf{Q}$ and of a polarizer
$\mathbf{L}$ oriented at an arbitrary angle are found by multiplying the Jones
matrices for the elements at $0$~degree
\cite{pedrotti} by rotation matrices:
\begin{equation}\label{eq:Q}
    \mathbf{Q}(\phi) = \frac{1}{\sqrt{2}}
    \left( \begin{array}{cc} 1 + i\cos(2\phi) & i\sin(2\phi)
    \\ i\sin(2\phi) & 1 - i\cos(2\phi)\end{array} \right) \quad ,
\end{equation}
\begin{equation}\label{eq:L}
    \mathbf{L}(\theta) =
    \frac{1}{2}\left( \begin{array}{cc} 1 + \cos(2\theta)& \sin(2\theta)
    \\ \sin(2\theta) & 1 - \cos(2\theta) \end{array} \right) \quad .
\end{equation}
The combined action of the elements in Eqs. (\ref{eq:pump}) to (\ref{eq:L}) on
the incoming wave $E_{in}$ now gives the electric field and
the intensity of the probe beam at the detector:
 \begin{equation}
\mathbf{E}_\mathrm{det}(\theta,\phi,\rho,\tau,\Delta\zeta)=\mathbf{L}(\theta)
\cdot \mathbf{Q}(\phi) \cdot \mathbf{D}(\rho,\tau) \cdot \mathbf{P}(\Delta\zeta)
\cdot \mathbf{E}_{in} \quad ,
\end{equation}
\begin{equation}\label{eq:intensity}
I_\mathrm{det}(\theta,\phi,\rho,\tau,\Delta\zeta)= \left| \mathbf{E}_\mathrm{det}
(\theta,\phi,\rho,\tau,\Delta\zeta) \right| ^2 \quad .
\end{equation}
A contour plot of the detected intensity $I_ \mathrm{det}$ as a function of $\theta$ and $\phi$
with $\rho=0$, $\tau=0$ and $\Delta\zeta=0$ is given in Fig.~\ref{wps_theory}.
The intensity has a minimum (dark fringe) for
$(\theta,\phi)=(45^\circ,45^\circ)$, increases outwards in an elliptical shape,
and at $(\theta,\phi)=(45^\circ,135^\circ)$ we find a bright fringe.

Essentially, with a lock-in amplifier, we subtract the detected intensity of the
probe beam when the pump is off from the detected intensity when the pump is on, which gives a signal of
\begin{equation}\label{eq:signal}
S(\theta,\phi,\rho,\tau,\Delta\zeta)=I_\mathrm{det}(\theta,\phi,\rho,\tau,
\Delta\zeta)-I_\mathrm{det}(\theta,\phi,\rho,\tau,0) \quad .
\end{equation}
In general, for a given value of $\theta$ and $\phi$, there will
be both a contribution from the real  and imaginary part  of
$\Delta\zeta$ to the signal in Eq.~(\ref{eq:signal}),
corresponding to amplitude and phase changes of the field,
respectively. They cannot be detected separately, except for
configurations where either the amplitude contribution or the
phase contribution is zero. For these working points, the detected
signal will consist of either the pure phase contribution, or the
pure amplitude contribution. In the following we call amplitude
working point a configuration of the interferometer which is
\emph{not} sensitive to a phase change in the field.  All these
amplitude working points form a line which we will call amplitude
line. It is the zero contour of the response to a phase change,
i.e. imaginary $\Delta\zeta$ in Eq.~(\ref{eq:signal}). In the same
way, a phase working point is a configuration of the
interferometer which is not sensitive to an amplitude change of
the field. The amplitude and phase lines are shown in
Fig.~\ref{wps_theory}.

Our model is fully analytical, and is in principle exact for a
weak enough probe pulse. It gives the measured signal in terms of
$\theta$, $\phi$, $\rho$, $\tau$, and $\Delta\zeta$, but this
expression is too lengthy for a qualitative discussion. Instead,
we will give a first-order expansion in $\theta$ and $\phi$ around
the dark fringe, which provides an illustrative explanation of the
working of the interferometer and of the separation of phase and
amplitude.

From Fig.~\ref{wps_theory}A we see that the interferometer is in
the dark fringe if the quarter-wave plate's fast axis and the
polarizer's transmission axis are both oriented at $+45$ degrees.
For a first order approximation, we only consider small deviations
from the dark fringe: $\hat{\theta} = \theta  - \frac{\pi}{4}$ and
$\hat{\phi} = \phi  - \frac{\pi}{4}$. Using the small-angle
approximation, neglecting all higher-order terms of $\hat{\theta}$
and $\hat{\phi}$, and disregarding the experimental correction
parameters $\rho$ and $\tau$, we find the total intensity at the
detector
\begin{equation}
I_\mathrm{det} = |\mathbf{E}_\mathrm{det}|^2 = \left(\hat{\theta} - \hat{\phi} -
\frac{\mathrm{Im}(\Delta\zeta)}{2}\right)^2 + \left( \hat{\phi} +
\frac{\mathrm{Re}(\Delta\zeta)}{2}\right)^2 \quad ,
\end{equation}
while the intensity at the detector without the pump
($\Delta\zeta=0$) is
\begin{equation}
I_{0} = (\hat{\theta} - \hat{\phi})^2 + \hat{\phi}^2 \quad .
\end{equation}
We thus find our signal, which is the first-order change in
intensity
\begin{equation}\label{eq:vii}
S =  I_\mathrm{det} - I_{0} = \hat{\phi} \  \mathrm{Re}(\Delta\zeta) -
(\hat{\theta} - \hat{\phi}) \ \mathrm{Im}(\Delta\zeta) \quad .
\end{equation}
From Eq.~(\ref{eq:vii}) it is easy to see that amplitude and phase
responses can be separated by choosing proper angles for the
polarizer and the quarter-wave plate. If $\hat{\theta}$ and
$\hat{\phi}$ are equal and nonzero, the interferometer is
sensitive for amplitude changes only, while if $\hat{\phi}=0$ and
$\hat{\theta} \neq 0$, the interferometer only measures phase
changes. Fig.~\ref{wps_theory}B shows that this  first-order
approximation still applies for the amplitude line even for large
angles, while the phase line is  described  by $\hat{\phi}=0$ only
very close to the dark fringe.

\section{Results and discussion}
\label{sec:results}

\subsection{Characterization of the interferometer}
\label{secsub:charac}

\begin{figure}[tbp]
\begin{center}
\includegraphics[width=120mm]{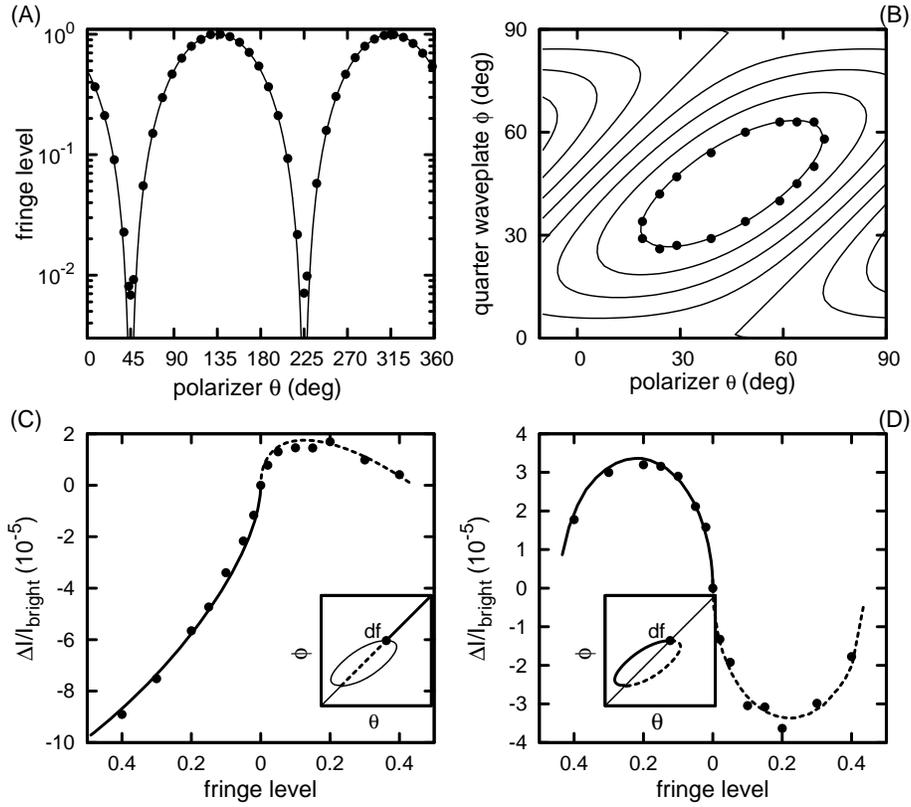}
\end{center}
\caption{\label{wps_data} (A) The contrast ratio of the
interferometer amounts to 150 as measured by scanning the
polarizer angle (dots). The model (line) assumes infinite
contrast, i.e. an absolute zero at the dark fringe. (B) The
position of the 10\%-fringe (dots) is well described by our model
(intensity contour lines in steps of 0.1). (C, D) Signal from a
single 60-nm gold nanoparticle, measured along the amplitude
working line (C) and along the phase working line (D). The lines
are the signals as calculated by the model assuming $\Delta \zeta
=  (-9.3 + 10.2 \, i) \, 10^{-5}$. The insets depict the
corresponding positions of the polarizer ($\theta$) and the
quarter-wave plate ($\phi$) relative to the dark fringe (df). The
measurement in panel (B) was carried out in the picosecond
configuration.}
\end{figure}

The contrast ratio of the interferometer ($I_{max}/I_{min}$),
found by rotating the polarizer  by 360 degrees as shown in
Fig.~\ref{wps_data}A, reaches 150 in the case shown here and
varies typically between 100 and 200. This corresponds to a fringe
visibility (defined as $(I_{max}-I_{min})/(I_{max}+I_{min})$)
between 98\% and 99\%. The contrast is mainly limited by
depolarization by the microscope objectives and by the quality of
the crystal surfaces. Nevertheless,  due to the simplified
alignment of the crystals, the fringe visibility is high compared
with other implementations of interferometric microscopes (90\%
ref. \cite{tachizaki06apr}, 66\% ref. \cite{hwang06feb}). As
expected, the model describes the position of the equal-intensity
contours, as shown in Fig.~\ref{wps_data}B for the 10\% contour
line.

The detection path was calibrated by  a laser beam
square-modulated at the lock-in reference frequency. In the
following we give, if not otherwise mentioned, the
root-mean-squared (rms) values of the pump-induced square
modulation to our probe beam. To be independent of the absolute
probe power, we normalize the intensity modulation $\Delta I$ to
the probe  intensity in the bright fringe $I_{bright}$.
Fig.~\ref{wps_data}C  and D compare the detected intensity
modulation  as a function of the interferometer fringe level for
working points on the amplitude  (Fig.~\ref{wps_data}C) and phase
line (Fig.~\ref{wps_data}D). The probe modulation was caused by a
single 60-nm gold particle in the focus under constant pump. The
data is well described by our model with a pump-induced field
modification of $\Delta \zeta =  (-9.3 + 10.2 \, i) \, 10^{-5}$ as
the only free parameter.


Note that the exact definition of $\Delta \zeta$ used in this
article differs from that used in our previous publication
\cite{vandijk05dec31}: we now give rms instead of peak-peak values
to facilitate comparison with noise levels and the change in the
probe field is given relative to the field at the sample and not
to the field at the detector. Both corrections reduce the absolute
value of  $\Delta \zeta$ by a factor of $2 \sqrt{2} \
I_{\mathrm{bright}} / I_{\mathrm{wp}}$. The   ratio
$I_{\mathrm{wp}}/ I_{\mathrm{bright}}$  used in
Ref.~\cite{vandijk05dec31} was given in the supplementary
material. In the linear-response regime it is possible to
calculate the pump-induced change of the probe field $\zeta(t)$
from $\Delta\zeta(t)$ (Eq.~\ref{eq:pump}). However, this
assumption is too crude for our experimental conditions, and the
full nonlinear response must be calculated for a quantitative
agreement between predicted and measured $\zeta(t)$ traces. This
calculation is however beyond the scope of this paper.

\subsection{Noise}
\label{secsub:noise}

\begin{figure}[tbp]
\begin{center}
\includegraphics{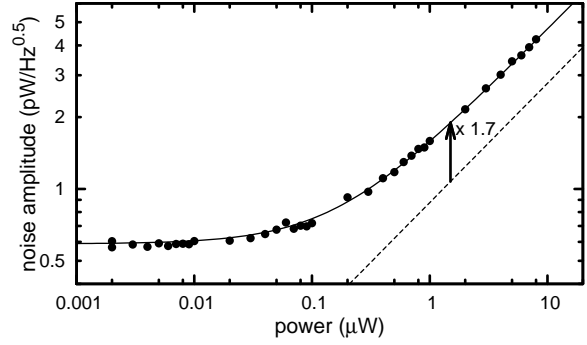}
\end{center}
\caption{\label{noise} Noise amplitude as a function of the
optical power at the detector. The solid line is a fit to
Eq.~(\ref{eq:noise3}), with $\sigma_{ampl}=0.6~
\mathrm{pW/\sqrt{Hz}}$, $n_{laser}=0$, and $F=3 \approx 1.7^2$.
The dashed line describes the expected photon noise scaling as the
square-root of the probe power. The difference is due to the
excess noise of the avalanche process in the analog detector. This
experiment was carried out in the picosecond configuration.}
\end{figure}

In an optical experiment, we have to consider two independent
sources of noise, electronic noise and optical shot noise, which
add quadratically. Optical noise itself has two causes. On the one
hand, laser noise $\sigma_{laser}$ stems from classical
fluctuations in the output power of the laser, and usually grows
linearly with laser power. The photon noise $\sigma_{ph}$, on the
other hand, originates from quantum-mechanical uncertainty in the
amplitude of the electric field of the laser wave and grows as the
square root of laser power. Unless squeezed light is used, photon
noise is the fundamental noise limit. Photon noise is given by
\cite{bachor_qo98}
\begin{equation}\label{eq:noise2}
    \sigma_{ph} = \sqrt{2 \; B \; P \; h\nu  } \quad ,
\end{equation}
where $B$ is the detection bandwidth and $P$ is the
detected optical power.

The electronic noise also has two dominant sources. One is the
thermal noise of the amplification stage of the detector
($\sigma_{ampl}$, dark noise), which is specified by the noise
equivalent power (NEP), and is independent of laser power. Another
source of electronic noise, the excess noise, is specific for the
analog avalanche photodiodes used here \cite{mcintyre66ieeet}. It
arises from statistical fluctuations in the avalanche process
following the absorption of every photon. Because it is related to
the detection of photons, excess noise has the same statistical
properties as photon noise and also increases as the square root
of laser power. The excess noise factor $F$ is defined as
$\sigma_{shot}^2=F \sigma_{ph}^2$, where  $\sigma_{shot}$ is the
full shot noise. The total noise amplitude $\sigma_{tot}$ then
amounts to

\begin{equation}\label{eq:noise3}
    \sigma_{tot}=\sqrt{ {\sigma_{ampl}}^2 +{\sigma_{shot}}^2  +{\sigma_{laser}}^2 } =
\sqrt{  {\sigma_{ampl}}^2 + F\; 2\; B \; P h\nu  + \left(n_{laser} P\right)^2}  \quad .
\end{equation}

The noise is measured as the standard deviation of a series of
calibrated root-mean-square values given by the lock-in amplifier.
This is equivalent to the standard deviation of a series of
absolute power measurements in the given bandwidth $B$.
Fig.~\ref{noise} shows the measured noise as a function of the
probe power at the detector (note that, as long as the pump laser
is sufficiently rejected from the detector with dichroic filters,
there is no pump-power dependence of the noise level). At low
power, the electronic noise dominates and we find a constant
contribution of 0.6~$\mathrm{pW/\sqrt{Hz}}$, within the
specification of the NEP of the detector
(section~\ref{sec:method}). At higher power, the optical noise
becomes dominant and in this regime, the noise scales as the
square-root of the laser power. The power range in
Fig.~\ref{noise} was limited by saturation of the detector. Within
this range, there is no contribution that scales linearly with
laser power, which means that laser noise does not contribute to
the total noise. For detected powers levels between 1 and
10~$\mu$W, the main noise component is shot noise. The expected
photon noise as a function of laser power is indicated with a
dashed line in Fig.~\ref{noise}.  The difference arises from the
excess noise generated by the avalanche photodiode and is
described by an excess noise factor $F=3 \approx 1.7^ 2$, in good
agreement with the manufacturer's specification. Replacing the
avalanche diode by a PIN diode will remove this excess-noise
contribution, if a sufficiently low-noise amplifier is used. The
analysis of noise sources in our experiment shows that the
interferometer is not absolutely required to reduce noise from
fluctuations of the light sources (laser noise), as our laser and
OPO systems are very stable. A much simpler experimental setup in
a pump-probe configuration would therefore suffice to provide
information on transient absorption only (i.e., on the amplitude
signal).

\subsection{Gold nanoparticles}
\label{secsub:gold}

As an illustration of some of the possibilities of our method, we
show experiments  on single gold nanoparticles. More exhaustive
results have been published elsewhere \cite{vandijk05dec31}. In
metal nanoparticles, optical contrast is mainly provided by the
surface plasmon-polariton, or Mie resonance
\cite{kreibig_vollmer}. The Mie resonance is a collective
oscillation of the free electrons in the particle, and manifests
itself as a relatively broad nearly-Lorentzian peak in the
absorption spectrum. The position and width of the peak depend on
the size and shape of the particle, and on its environment
\cite{kreibig_vollmer}. If a particle is excited by a short
near-infrared laser pulse, first only free electrons are excited.
By electron-electron scattering, a thermalized electron gas is
quickly formed, which leads to a short but strong increase in the
electron pressure \cite{perner00jul24} and a broadening of the
plasmon resonance \cite{perner97mar17}. The electrons subsequently
cool by electron-phonon coupling on a timescale of about 1~ps,
hereby heating the lattice. Note that this cooling time depends on
the intensity of the excitation \cite{hartland04physc}. This
sudden heating of the lattice, together with the short peak of
electronic Fermi pressure immediately after absorption of the
light pulse, launches an acoustic vibration in the lattice
\cite{perner00jul24}, which manifests itself optically as a
periodic red shift of the plasmon resonance
\cite{hartland04physc,delfatti99jun15}. The period of this radial
breathing mode is given in Lamb's theory \cite{lamb1882} as a
function of the particle size and of the longitudinal and
transverse sound velocities of the metal \cite{hodak99nov8}. The
period  is on the order of 20 picoseconds for free gold spheres of
about 60 nm diameter. On a longer timescale (100-1000~ps), the
lattice cools through heat conduction to the environment.

\begin{figure}[tbp]
\begin{center}
\includegraphics[width=120mm]{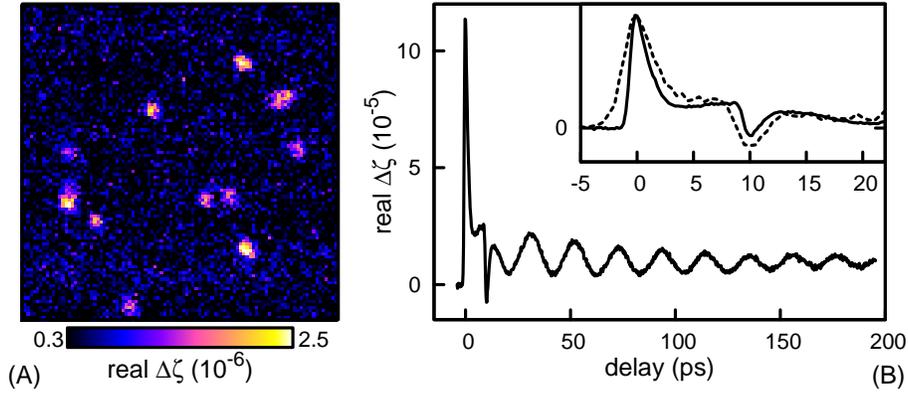}
\end{center}
\caption{\label{gold} (A) Confocal scan of a $10 \ \mathrm{\mu m}
\times  10 \ \mathrm{\mu m}$ area of a sample containing 10-nm
gold nanoparticles. Each diffraction limited spot is caused by a
single nanoparticle. The background contains only the noise
discussed in section \ref{secsub:noise}. Probe wavelength and
pump-probe delay were optimized for maximum contrast (532 nm, 0
ps, respectively). This experiment was carried out in the
picosecond configuration. The sampling time was 200~ms per pixel.
(B) Example of a delay scan of a single 60-nm gold nanoparticle
showing a short spike due to the hot electron gas and periodic
oscillations of the particle size. The inset compare the
'femtosecond' (solid) and the 'picosecond' (dashed) configuration
of the laser system in their resolving power of the fast
electronic process. The traces in the inset are normalized to the
first peak. The detection bandwidth was 7.8~Hz.}
\end{figure}

In Fig.~\ref{gold}A, a confocal scan of a sample containing 10-nm
gold nanoparticles is shown. Each diffraction-limited spot is
caused by a single nanoparticle \cite{vandijk05dec31}. The
background consists only of the noise discussed in section
\ref{secsub:noise}. The delay between the pump and the
probe-reference pair was tuned such that the pump and the probe
arrive at the sample simultaneously. In this way, we measure the
prompt response of the nanoparticle to the excitation, which gives
the highest contrast on the real part of the signal (absorptive
response). The wavelength of the probe was tuned to the plasmon
resonance of the particles at 532~nm, again to ensure maximum
contrast. With a sampling time of 200~ms per pixel, the particles
are detected with a signal-to-noise ratio of roughly 7. Pump and
probe power are limited by the maximum permissible absorption in
the particle, which results in a detection limit of 10-nm diameter
for gold nanoparticles, for reasonable sampling times shorter than
one second per pixel.

We measure the time-dependent response of the nanoparticle to
excitation with the pump pulse by varying the delay between the
pump pulse and the probe-reference pair of pulses.
Fig.~\ref{gold}B shows an example of the response of a single
60-nm gold nanoparticle in the laser focus, detected with a
bandwidth of 7.8~Hz. The first part of the trace, which is
enlarged in the inset, shows two sharp peaks, that occur when the
pump is overlapped with the probe and with the reference
respectively. These peaks stem from the change in optical response
due to the hot electron gas. From the rise and decay times of the
peaks, the electron-electron scattering and electron-phonon
coupling times can be obtained \cite{hartland04physc,voisin04may}.
Since these processes take place on timescales faster than a
picosecond, the width of the peaks is dominated by the pulse
length if the measurement is carried out with picosecond pulses,
and information on the electronic decay times of the particle is
lost. However, by using shorter pulses, it is possible to resolve
these processes. This is demonstrated in the inset of
Fig.~\ref{gold}B, where the fast electronic response is measured
with femtosecond and picosecond pulses on two different single
particles. In the femtosecond experiment, the peaks are narrow
enough to resolve the electronic processes in the particle. On a
longer timescale, the signal shows a damped harmonic oscillation.
This is a direct observation of the acoustic vibrations of the
particle, from which we can derive information on the elastic
properties of the particle itself and its mechanical coupling to
the surroundings.

\begin{figure}[tbp]
\begin{center}
\includegraphics[width=120mm]{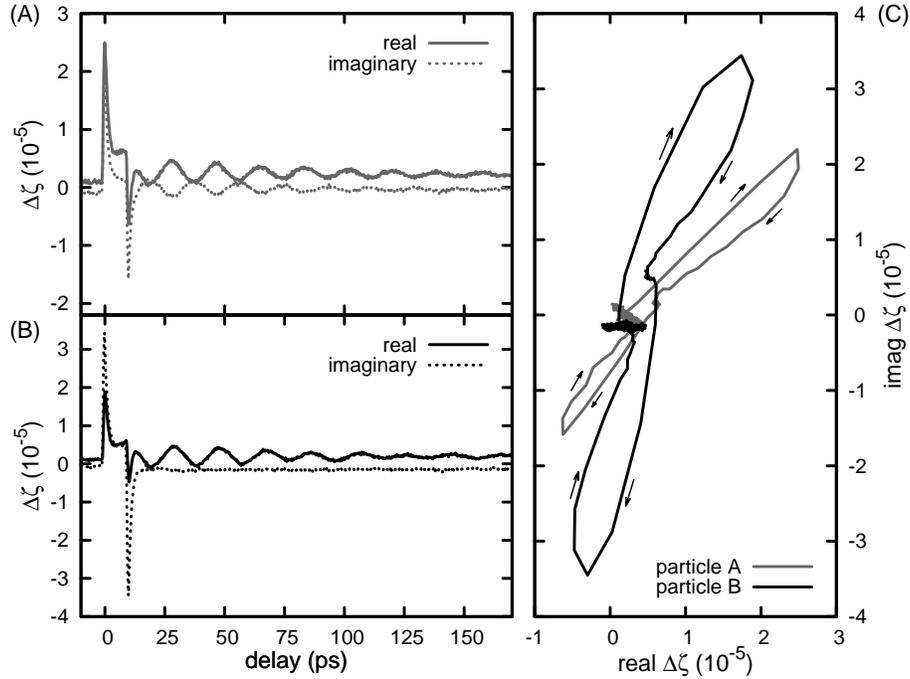}
\end{center}
\caption{\label{fullresp} (A,B) Delay scans of  two gold
nanoparticles with a nominal diameter of 60 nm, measured at the
amplitude (solid) and phase (dotted) sensitive working point. For
particle (A), the peaks have the same sign, but the sign of the
vibrations is opposite demonstrating the different spectral
origin. For  the particle in panel (B), the dispersive
contribution to the peaks is much stronger, while the vibrations
completely vanish in this phase-sensitive configuration. (C)
Plotting Im($\Delta\zeta$) against Re($\Delta\zeta$) reveals the
full response of both particles in the complex plane. The arrows
indicate the time evolution. In all plots, the detected intensity
was 2\% of the bright-fringe intensity. The traces show the
average of five measurements, each with a detection bandwidth of
7.8~Hz. }
\end{figure}

In Fig.~\ref{fullresp}, we combine the amplitude and phase
separation with time-resolved experiments, in order to measure the
complete complex temporal response of the particle. As mentioned
in section~\ref{sec:model}, for delay times larger than 10~ps, it
is important to choose a working point close to the dark fringe,
to ensure the full separation of amplitude and phase. The
interferometer was tuned to a 2\%~fringe, to comply with the
approximation of Eq.~(\ref{eq:pump}). In Fig.~\ref{fullresp}A, the
real and imaginary temporal responses of a single particle are
shown together. Here, the different spectral origin (resonance
broadening versus red shift) of the peaks and of the oscillations
becomes apparent, since the two peaks have the same sign in both
the amplitude and the phase configuration, while the vibrations
are exactly out of phase. Figure~\ref{fullresp}B shows the real
and imaginary temporal response of another particle in the same
sample. While the amplitude response of this particle is nearly
the same as that of the particle in Fig~\ref{fullresp}A, the phase
response is rather different. The electronic contribution gives a
mainly dispersive response, while the acoustic vibrations are
completely absorptive. The differences between the two particles
can be visualized in a different manner in Fig.~\ref{fullresp}C,
where the dispersive response is plotted against the absorptive
response for both particles. The trace of the particle from
Fig.~\ref{fullresp}A (gray) shows a weaker electronic response,
while the overall trace is tilted towards the real axis. These
kind of differences between particles can arise from differences
in their spectral responses, themselves due to differences in
their size, shape or local environment. Further study is needed to
precisely correlate the optical and structural properties of the
particles, but there is little doubt that single-particle studies
of gold nanoparticles will lead to a better understanding of their
properties, and to such applications as elasticity and damping
sensors at nanometer scales.

\section{Conclusion}

We have developed a common-path interferometer, which enables
time-resolved experiments on single nanoparticles, and with which
the absorptive and dispersive responses of the particles can be
separated. The design of the interferometer, using birefringent
crystals as splitting elements, simplifies the alignment and
increases the fringe contrast. A noise analysis showed that the
detection is shot-noise limited, just a factor of about 1.7 above
the ultimate limit of photon noise. We expect to reach the
photon-noise limit by replacing the avalanche diode with a good
PIN diode. The interferometer is fully described by an analytical
model, which allows us to find optimal working points. We have
successfully used our method for time-resolved experiments on
single gold nanoparticles. While also direct absorption
measurements on single particles yield the ultrafast response of
the hot electron gas \cite{muskens06mar}, only an interferometric
experiment provides the full complex response of a nanoparticle on
the sub-picosecond timescale, free from any model or assumption.

M.L.\ acknowledges a Marie Curie Fellowship from the European
Commission (Contract no.\ HPMF-CT-2002-02099). This work is part
of the research program of the "Stichting voor Fundamenteel
Onderzoek der Materie" (FOM), financially supported by the NWO.


\begin{thebibliography}{10}
\newcommand{\enquote}[1]{``#1''}
\expandafter\ifx\csname url\endcsname\relax
  \def\url#1{\texttt{#1}}\fi
\expandafter\ifx\csname
urlprefix\endcsname\relax\def\urlprefix{URL }\fi
\providecommand{\eprint}[2][]{\url{#2}}

\bibitem{science99}
{Special issue on Single Molecules}, Science \textbf{283}(5408),
1593--1804
  (1999).

\bibitem{vandijk05jul}
M.~A. {van Dijk}, M.~Lippitz, and M.~Orrit, \enquote{Far-field
optical
  microscopy of single metal manoparticies,} Accounts Chem. Res.
  \textbf{38}(7), 594--601 (2005).

\bibitem{vandijkpccp}
M.~A. {van Dijk}, A.~L. Tchebotareva, M.~Orrit, M.~Lippitz,
S.~Berciaud,
  D.~Lasne, L.~Cognet, and B.~Lounis, \enquote{Absorption and scattering
  microscopy of single metal nanoparticles,} Phys. Chem. Chem. Phys.
  \textbf{8}, 3486 -- 3495 (2006).

\bibitem{cognet03sep30}
L.~Cognet, C.~Tardin, D.~Boyer, D.~Choquet, P.~Tamarat, and
B.~Lounis,
  \enquote{Single metallic nanoparticle imaging for protein detection in
  cells,} Proc. Natl. Acad. Sci. U. S. A. \textbf{100}(20), 11,350--11,355
  (2003).

\bibitem{vandijk05dec31}
M.~A. {van Dijk}, M.~Lippitz, and M.~Orrit, \enquote{Detection of
acoustic
  oscillations of single gold nanospheres by time-resolved interferometry,}
  Phys. Rev. Lett. \textbf{95}(26), 267,406 (2005).

\bibitem{jin05sep14}
R.~C. Jin, J.~E. Jureller, H.~Y. Kim, and N.~F. Scherer,
\enquote{Correlating
  second harmonic optical responses of single Ag nanoparticles with
  morphology,} J. Am. Chem. Soc. \textbf{127}(36), 12,482--12,483 (2005).

\bibitem{berciaud05mar}
S.~Berciaud, L.~Cognet, P.~Tamarat, and B.~Lounis,
\enquote{Observation of
  intrinsic size effects in the optical response of individual gold
  nanoparticles,} Nano Lett. \textbf{5}(3), 515--518 (2005).

\bibitem{muskens06mar}
O.~L. Muskens, N.~{Del Fatti}, and F.~Vall{\'e}e,
\enquote{Femtosecond response
  of a single metal nanoparticle,} Nano Lett. \textbf{6}(3), 552--556 (2006).

\bibitem{lippitz05apr}
M.~Lippitz, M.~A. {van Dijk}, and M.~Orrit,
\enquote{Third-harmonic generation
  from single gold nanoparticles,} Nano Lett. \textbf{5}(4), 799--802 (2005).

\bibitem{stoller06aug15}
P.~Stoller, V.~Jacobsen, and V.~Sandoghdar, \enquote{Measurement
of the complex
  dielectric constant of a single gold nanoparticle,} Opt. Lett.
  \textbf{31}(16), 2474--2476 (2006).

\bibitem{arbouet04sep17}
A.~Arbouet, D.~Christofilos, N.~{Del Fatti}, F.~Vall{\'e}e, J.~R.
Huntzinger,
  L.~Arnaud, P.~Billaud, and M.~Broyer, \enquote{Direct measurement of the
  single-metal-cluster optical absorption,} Phys. Rev. Lett. \textbf{93}(12),
  127,401 (2004).

\bibitem{lindfors04jul16}
K.~Lindfors, T.~Kalkbrenner, P.~Stoller, and V.~Sandoghdar,
\enquote{Detection
  and spectroscopy of gold nanoparticles using supercontinuum white light
  confocal microscopy,} Phys. Rev. Lett. \textbf{93}(3), 037,401 (2004).

\bibitem{ignatovich06jan13}
F.~V. Ignatovich and L.~Novotny, \enquote{Real-time and
background-free
  detection of nanoscale particles,} Phys. Rev. Lett. \textbf{96}(1), 013,901
  (2006).

\bibitem{boyer02aug16}
D.~Boyer, P.~Tamarat, A.~Maali, B.~Lounis, and M.~Orrit,
\enquote{Photothermal
  imaging of nanometer-sized metal particles among scatterers,} Science
  \textbf{297}(5584), 1160--1163 (2002).

\bibitem{berciaud04dec17}
S.~Berciaud, L.~Cognet, G.~A. Blab, and B.~Lounis,
\enquote{Photothermal
  heterodyne imaging of individual nonfluorescent nanoclusters and
  nanocrystals,} Phys. Rev. Lett. \textbf{93}(25), 257,402 (2004).

\bibitem{kop95dec}
R.~H.~J. Kop and R.~Sprik, \enquote{Phase-sensitive interferometry
with
  ultrashort optical pulses,} Rev. Sci. Instrum. \textbf{66}(12), 5459--5463
  (1995).

\bibitem{zernike42}
F.~Zernike, \enquote{Phase contrast, a new method for the
microscopic
  observation of transparent objects,} Physica \textbf{9}, Part I, 686--698,
  Part II, 974--986 (1942).

\bibitem{hwang06feb}
J.~Hwang, M.~M. Fejer, and W.~E. Moerner, \enquote{Scanning
interferometric
  microscopy for the detection of ultrasmall phase shifts in condensed matter,}
  Phys. Rev. A \textbf{73}(2), 021,802 (2006).

\bibitem{lagasse89mar15}
M.~J. LaGasse, D.~Liu-Wong, J.~G. Fujimoto, and H.~A. Haus,
\enquote{Ultrafast
  switching with a single-fiber interferometer,} Opt. Lett. \textbf{14}(6),
  311--313 (1989).

\bibitem{patel98may10}
N.~S. Patel, K.~L. Hall, and K.~A. Rauschenbach,
\enquote{Interferometric
  all-optical switches for ultrafast signal processing,} Appl. Optics
  \textbf{37}(14), 2831--2842 (1998).

\bibitem{hurley99sep15}
D.~H. Hurley and O.~B. Wright, \enquote{Detection of ultrafast
phenomena by use
  of a modified Sagnac interferometer,} Opt. Lett. \textbf{24}(18), 1305--1307
  (1999).

\bibitem{sugawara02may6}
Y.~Sugawara, O.~B. Wright, O.~Matsuda, M.~Takigahira, Y.~Tanaka,
S.~Tamura, and
  V.~E. Gusev, \enquote{Watching ripples on crystals,} Phys. Rev. Lett.
  \textbf{88}(18), 185,504 (2002).

\bibitem{tachizaki06apr}
T.~Tachizaki, T.~Muroya, O.~Matsuda, Y.~Sugawara, D.~H. Hurley,
and O.~B.
  Wright, \enquote{Scanning ultrafast Sagnac interferometry for imaging
  two-dimensional surface wave propagation,} Rev. Sci. Instrum. \textbf{77}(4),
  043,713 (2006).

\bibitem{pedrotti}
F.~L. Pedrotti and L.~S. Pedrotti, \emph{Introduction to Optics}
(Prentice
  Hall, 1993).

\bibitem{bachor_qo98}
H.-A. Bachor, \emph{A Guide to Experiments in Quantum Optics}
(Wiley-VCH,
  1998).

\bibitem{mcintyre66ieeet}
R.~J. McIntyre, \enquote{Multiplication Noise In Uniform Avalanche
Diodes,}
  IEEE Trans. Electron Devices \textbf{ED13}(1), 164--168 (1966).

\bibitem{kreibig_vollmer}
U.~Kreibig and M.~Vollmer, \emph{Optical Properties of Metal
Clusters}, vol.~25
  of \emph{Springer Series in Materials Science} (Springer, Berlin, 1995).

\bibitem{perner00jul24}
M.~Perner, S.~Gresillon, J.~Marz, G.~{von Plessen}, J.~Feldmann,
  J.~Porstendorfer, K.~J. Berg, and G.~Berg, \enquote{Observation of
  hot-electron pressure in the vibration dynamics of metalnanoparticles,} Phys.
  Rev. Lett. \textbf{85}(4), 792--795 (2000).

\bibitem{perner97mar17}
M.~Perner, P.~Bost, U.~Lemmer, G.~von Plessen, J.~Feldmann,
U.~Becker,
  M.~Mennig, M.~Schmitt, and H.~Schmidt, \enquote{Optically induced damping of
  the surface plasmon resonance in gold colloids,} Phys. Rev. Lett.
  \textbf{78}(11), 2192--2195 (1997).

\bibitem{hartland04physc}
G.~V. Hartland, \enquote{Measurements of the material properties
of metal
  nanoparticles by time-resolved spectroscopy,} Phys. Chem. Chem. Phys.
  \textbf{6}(23), 5263--5274 (2004).

\bibitem{delfatti99jun15}
N.~{Del Fatti}, C.~Voisin, F.~Chevy, F.~Vall{\'e}e, and
C.~Flytzanis,
  \enquote{Coherent acoustic mode oscillation and damping in silver
  nanoparticles,} J. Chem. Phys. \textbf{110}(23), 11,484--11,487 (1999).

\bibitem{lamb1882}
H.~Lamb, \enquote{On the Vibrations of an Elastic Sphere,}
Proceedings of the
  London Mathematical Society \textbf{13}, 189--212 (1882).

\bibitem{hodak99nov8}
J.~H. Hodak, A.~Henglein, and G.~V. Hartland, \enquote{Size
dependent
  properties of Au particles: Coherent excitation and dephasing of acoustic
  vibrational modes,} J. Chem. Phys. \textbf{111}(18), 8613--8621 (1999).

\bibitem{voisin04may}
C.~Voisin, D.~Christofilos, P.~A. Loukakos, N.~Del~Fatti,
F.~Vall{\'e}e,
  J.~Lerme, M.~Gaudry, E.~Cottancin, M.~Pellarin, and M.~Broyer,
  \enquote{Ultrafast electron-electron scattering and energy exchanges in
  noble-metal nanoparticles,} Phys. Rev. B \textbf{69}(19), 195,416 (2004).

\end{thebibliography}
\end{document}